\documentclass[iop,revtex4]{emulateapj}

\usepackage{aas_macros}
\usepackage{natbib}
\usepackage{array}
\usepackage{graphicx}
\usepackage{latexsym}
\usepackage{amsfonts,amsmath,amssymb}
\usepackage{url}
\usepackage[utf8]{inputenc}
\usepackage{hyperref}
\usepackage{placeins}
\hypersetup{colorlinks=false,pdfborder={0 0 0},}

\shortauthors{Jean-Luc Margot}

\begin{document}

\title{A Quantitative Criterion for Defining Planets}

\author{Jean-Luc Margot}

\affil{University California, Los Angeles}

\begin{abstract}
A simple metric can be used to determine whether a planet or exoplanet
can clear its orbital zone during a characteristic time scale, such as
the lifetime of the host star on the main sequence.  This criterion
requires only estimates of star mass, planet mass, and orbital period,
making it possible to immediately classify 99\% of all known
exoplanets.  All 8 planets and all classifiable exoplanets satisfy the
criterion.  This metric may be useful in generalizing and simplifying
the definition of a planet.
\end{abstract}

\bibliographystyle{apj}

\section{Introduction}
\label{sec-intro}
In 2006, the International Astronomical Union (IAU) adopted resolution
B5, which states: ``A planet is a celestial body that (a) is in orbit
around the Sun, (b) has sufficient mass for its self-gravity to
overcome rigid body forces so that it assumes a hydrostatic
equilibrium (nearly round) shape, and (c) has cleared the
neighbourhood around its orbit.''

Here, we propose a simple metric that allows for the quantification of
the third requirement and the extension of the definition to planets
orbiting other stars.

It must be emphasized at the outset that a planet can never completely
clear its orbital zone, because gravitational and radiative forces
continually perturb the orbits of asteroids and comets into
planet-crossing orbits.  What the IAU intended is not the impossible
standard of impeccable orbit clearing; rather the standard is
analogous to what \citet{sote06,sote08} described as a
{\em dynamical-dominance} criterion.  In this article, we use the IAU
orbit-clearing language even though the dynamical-dominance language
seems less prone to misinterpretation.

\section{Existing metric}
\label{sec-metric}
We seek to determine whether a celestial body can clear the
neighbourhood around its orbit.  To do so, we adapt the criterion that
\citet{trem93} developed for the formation of Oort-type comet clouds.
He considered the ejection of comets by a single planet of mass $M_p$
on a circular orbit of radius $a_p$ around a central star of mass
$M_\star$.  The ejection process is a diffusion or random walk process
in the comet's orbital energy, with the energy described by the
variable $x = 1/a$ where $a$ is the semi-major axis of the comet's
orbit.  The diffusion coefficient $D_x = \langle(\Delta
x)^2\rangle^{1/2}$ is the root mean square change in $x$ per orbit
resulting from gravity kicks from the planet.  Based on earlier work
by \citet{fern81} and \citet{dunc87}, \citet{trem93} found
\begin{equation}
D_x \simeq \frac{10}{a_p}\frac{M_p}{M_\star}.
\label{eq-Dx}
\end{equation}
The characteristic number of orbits required for the energy to change
by an amount equal to itself is $(x^2/D_x^2)$ and the corresponding
diffusion time is $t_{\rm diff} = P(x^2/D_x^2)$, where $P$ is the
comet's orbital period.  \citet{trem93} considered comets on orbits
initially similar to the orbit of the planet, i.e., $x = a_p^{-1}$, and
computed the planet mass required for the comet diffusion time to be
less than the age of the planetary system $t_\star$.  He found
\begin{equation}
\frac{M_p}{M_\earth} \gtrsim \left(\frac{M_\star}{M_\sun}\right)^{3/4} \left(\frac{t_\star}{10^{9}\, {\rm y}}\right)^{-1/2} \left(\frac{a_p}{1\, {\rm au}}\right)^{3/4},
\end{equation}
where the symbols $\earth$ and $\sun$ refer to Earth and Sun,
respectively.  This criterion has the same dependence as the criterion
$\Lambda \propto M_p^2/P$ that has been considered by
others~\citep{ster02,sote06}.

\section{Proposed metric}
By requiring a change in energy equal to the initial orbital energy,
\citet{trem93} constructed a condition that ensures that small bodies
are scattered out to very large distances, which is the proper
criterion when contemplating the formation of Oort-type comet clouds.
Here, we are interested in a criterion that meets the third
requirement of the 2006 IAU planet definition, i.e., that a planet clears
its orbital zone.  Ejection is not required.  Instead, what is needed
is a change in orbital energy that is sufficient to evacuate the small
bodies out to a certain distance $CR_H$, where $C$ is a numerical
constant and $R_H$ is the Hill radius of the planet:
\begin{equation}
R_{H} = \left( \frac{M_p}{3M_\star} \right)^{1/3}a_p.
\label{eq-rh}
\end{equation}
The value of $C$ must exceed $2\sqrt{3}$ to ensure that the planet
clears its feeding zone~\citep{birn73,arty87,ida93,glad93}.  A more stringent
condition would impose clearing a zone extending to 5 Hill radii
($C=5$).  The latter value mirrors certain stability criteria and the
observed dynamical spacing between exoplanets
(Section~\ref{sec-discuss}).

Consider small bodies initially on orbits similar to the orbit of the
planet, with $a \simeq a_p$.  The required energy change for clearing
a region of size $CR_H$ around the orbit is  
\begin{equation}
\delta x = \frac{1}{a_p} - \frac{1}{a_p + CR_H} = \frac{CR_H/a_p}{a_p(1 + CR_H/a_p)}.
\end{equation}
Ignoring the second term in the denominator provides a lower bound on
the energy requirement and an excellent approximation in most situations:
\begin{equation}
\delta x \simeq \frac{C}{a_p}\left(\frac{M_p}{3M_\star}\right)^{1/3}.
\label{eq-deltax}
\end{equation}
We can define the clearing time as
\begin{equation}
t_{\rm clear} = P \frac{\delta x^2}{D_x^2},
\end{equation}
and use equations~(\ref{eq-Dx}) and (\ref{eq-deltax}) with the orbital
period $P=2\pi a_p^{3/2}/(GM_\star)^{1/2}$ to arrive at
\begin{equation}
t_{\rm clear} = C^2 1.1\times10^5\, {\rm y} \left(\frac{M_\star}{M_\sun}\right)^{5/6} \left(\frac{M_p}{M_\earth}\right)^{-4/3} \left(\frac{a_p}{1\, {\rm au}}\right)^{3/2}.
\end{equation}
An Earth-mass planet orbiting a solar-mass star
at 1 au can clear its orbital zone to $2\sqrt{3}$ Hill radii in $\sim 1$ My.

\begin{figure}[h]
  \includegraphics[width=\columnwidth]{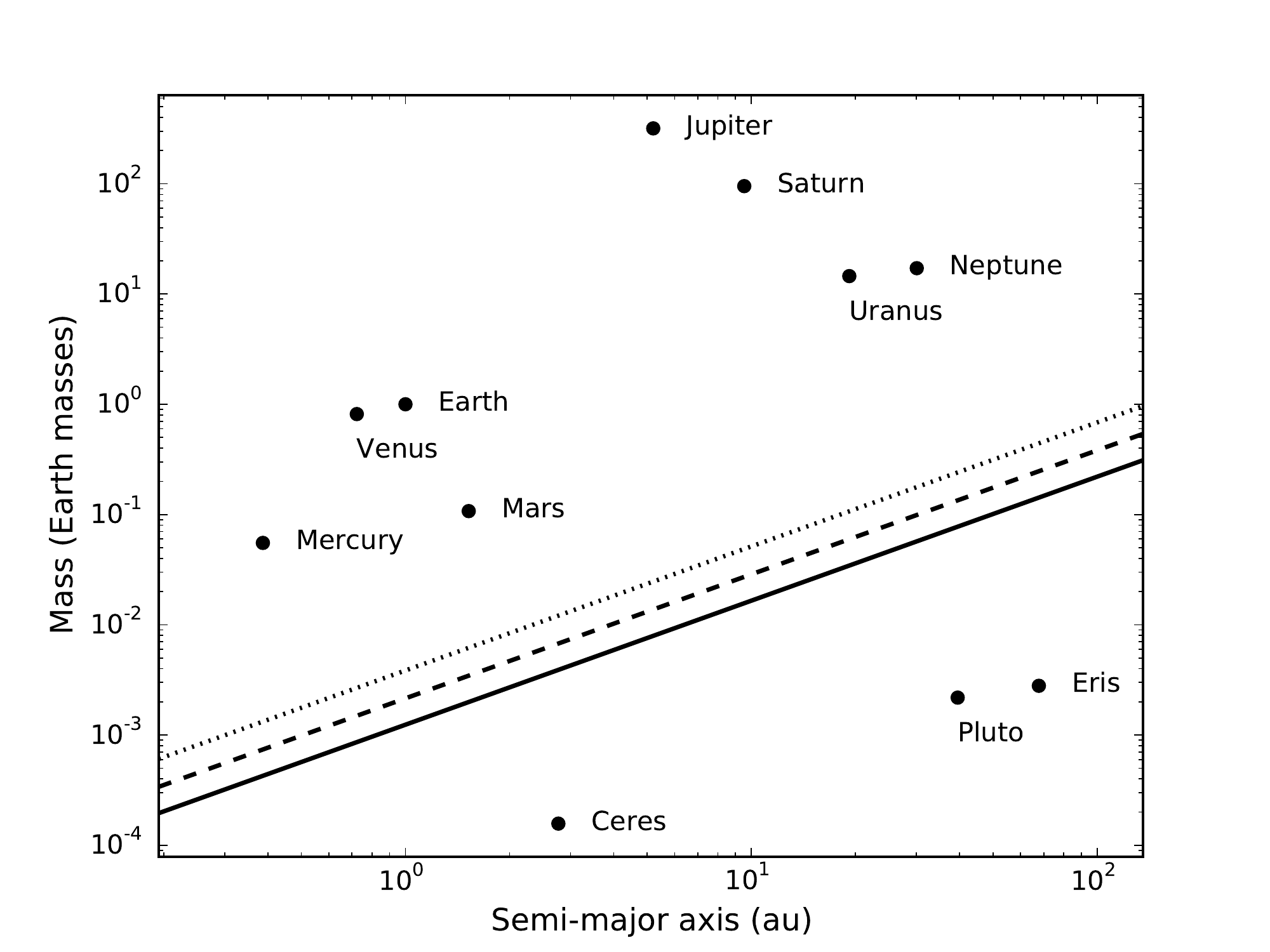}
\caption{Mass required to clear an orbital zone as a function of
  semi-major axis for a host star of mass 1 $M_\sun$.  The top two
  lines show clearing to 5 Hill radii in either 10 billion years
  (dashed line) or 4.6 billion years (dotted line).  The solid line
  shows clearing of the feeding zone ($2\sqrt{3}$ Hill radii) in 10
  billion years.}
\label{fig-ss}
\end{figure}

\begin{figure}[h]
  \includegraphics[width=\columnwidth]{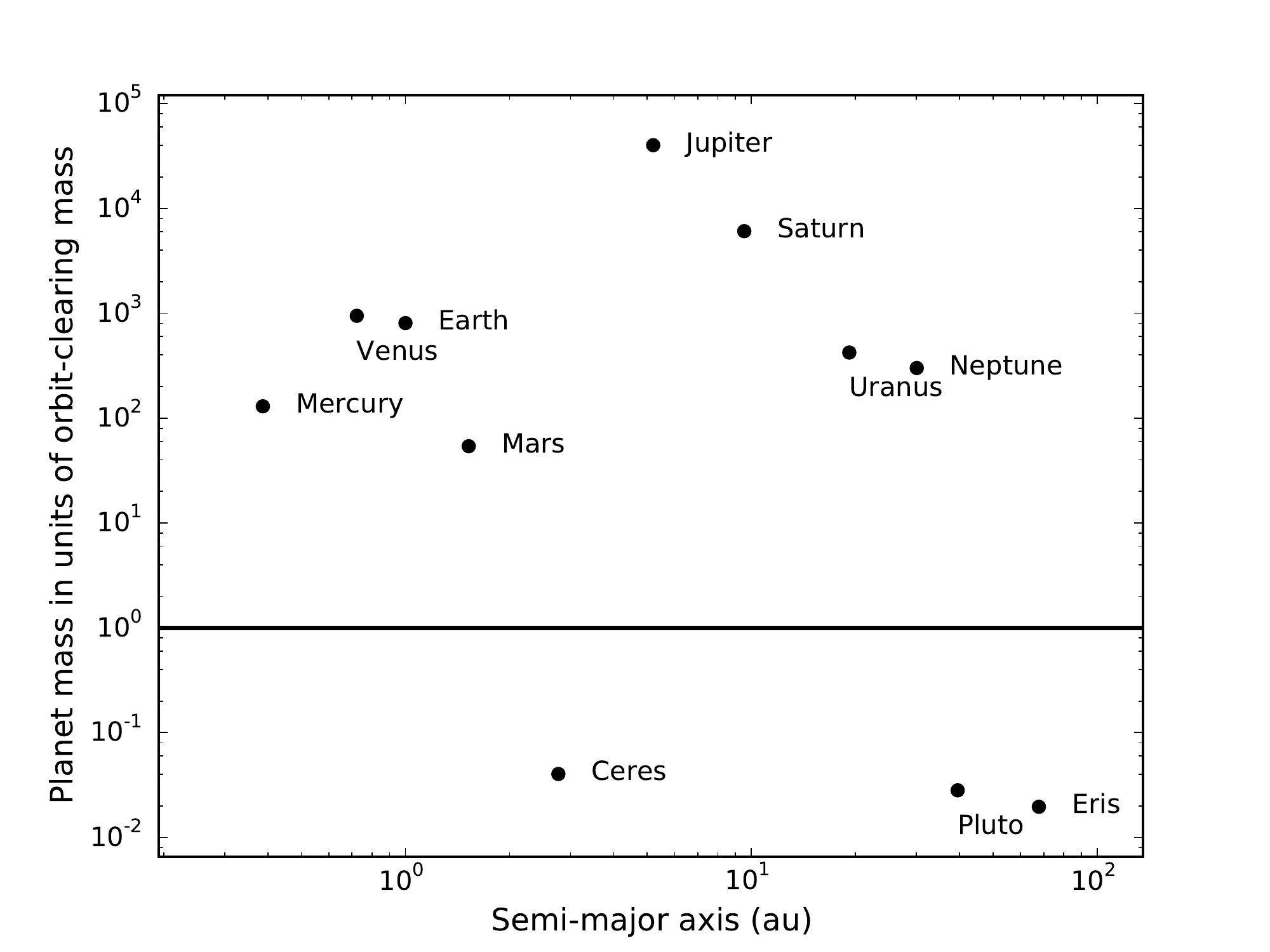}
  \caption{Planet test ($C=2\sqrt{3}, t_\star=t_{\rm MS}$) applied to objects in
    the solar system.  All 8 planets have a mass that exceeds the mass
    required to clear the corresponding orbital zone ($\Pi \geq 1$).}
\label{fig-testss}
\end{figure}

In the spirit of the IAU resolution, we suggest that a body that is
capable of clearing its orbit within a well-defined time interval is a
planet.  Requiring that the clearing time be less than $t_\star$,
which is now understood as a characteristic time related to the host
star, we find an expression for the minimum orbit-clearing mass:
\begin{equation}
\frac{M_p}{M_\earth} \gtrsim C^{3/2} \left(\frac{M_\star}{M_\sun}\right)^{5/8} \left(\frac{t_\star}{ 1.1\times10^5\, {\rm y}}\right)^{-3/4} \left(\frac{a_p}{1\, {\rm au}}\right)^{9/8}.
\label{eq-ss}
\end{equation}
This relationship clearly distinguishes the 8 planets in the solar
system from all other bodies (Figures~\ref{fig-ss} and \ref{fig-testss}).

For main sequence stars, a sensible characteristic time scale is the
host star's lifetime on the main sequence, i.e., $t_\star \simeq t_{\rm
  MS}$.  Incorporating the approximate relationship $t_{\rm MS}/t_\sun
\propto (M_\star/M_\sun)^{-2.5}$ with $t_\sun = 10^{10}$ y into equation
(\ref{eq-ss}), we find
\begin{equation}
  \frac{M_p}{M_\earth} \gtrsim 1.9 \times 10^{-4} C^{3/2} \left(\frac{M_\star}{M_\sun}\right)^{5/2} \left(\frac{a_p}{1\, {\rm au}}\right)^{9/8}.
\label{eq-all}
\end{equation}
For most stars of interest, the main sequence lifetime has
uncertainties up to a factor of 2 and the corresponding uncertainty on
the orbit-clearing mass is $<2$.

We use the notation $M_{\rm clear}$ to represent the orbit-clearing
mass given by the right-hand side of equation (\ref{eq-ss}) or
(\ref{eq-all}) and we use the symbol $\Pi$ to represent the mass of a
planetary body in terms of the corresponding orbit-clearing mass:
\begin{equation}
\Pi = \frac{M_{\rm body}}{M_{\rm clear}}.
\end{equation}
A simple planet test consists of evaluating whether the discriminant
$\Pi$ exceeds 1.  Values of $\Pi$ for solar system bodies are listed
in Table~\ref{tab-ss} and shown in Figure \ref{fig-testss}.

\begin{table}[b]
  \begin{center}
    \begin{tabular}{lrr}
      Body & \multicolumn{1}{c}{Mass ($M_\earth$)} & \multicolumn{1}{c}{$\Pi$}\\
  \\[-2mm]
  \hline \\[-2mm]
   Jupiter &     317.90 &      $4.0 \times 10^{4}$   \\
    Saturn &      95.19 &      $6.1 \times 10^{3}$   \\
     Venus &       0.815 &      $9.5 \times 10^{2}$  \\
     Earth &       1.000 &      $8.1 \times 10^{2}$  \\
    Uranus &      14.54 &      $4.2 \times 10^{2}$   \\
   Neptune &      17.15 &      $3.0 \times 10^{2}$   \\
   Mercury &       0.055 &      $1.3 \times 10^{2}$  \\
      Mars &       0.107 &      $5.4 \times 10^{1}$  \\
      \\[-2mm]
      \hline \\[-2mm]
     Ceres &       $1.6 \times 10^{-4}$ &     $4.0 \times 10^{-2}$   \\
     Pluto &       $2.2 \times 10^{-3}$ &     $2.8 \times 10^{-2}$   \\
      Eris &       $2.8 \times 10^{-3}$ &     $2.0 \times 10^{-2}$   \\
      \\[-2mm]
      \hline \\[-2mm]
\end{tabular}
    \caption{Values of the planet discriminant $\Pi$ $(C=2\sqrt{3}, t_\star=t_{\rm MS}$) for solar system bodies.}
    \label{tab-ss}
  \end{center}
\end{table}

The proposed metric for classifying planets is attractive because it
relies solely on properties that are typically known (i.e., host star
mass) or observable from Earth shortly after discovery (i.e., planet
mass and semi-major axis or orbital period).  When the planet mass is
not directly available, other observables (e.g., planet radius) can be
used to place bounds on $M_p$.  In the next section, we use equation
(\ref{eq-ss}) or (\ref{eq-all}) to test whether known exoplanets can
clear their orbits.

\clearpage
\enlargethispage{.5cm}
\section{Classification of exoplanets}

We applied the proposed planet criterion to the exoplanets listed in
the NASA Exoplanet
Archive.\footnote{http://exoplanetarchive.ipac.caltech.edu}  We were
able to classify 4620/4664 (99\%) of Kepler objects, 829/849 (98\%) of
non-Kepler objects, and 5/5 (100\%) of pulsar objects.

\subsection{Kepler objects}
\label{sec-koi}
As of July 17, 2015, the archive contained 4664 Kepler Objects of
Interest (KOI) that were not labeled as false positive.  Of those,
1001 were marked as confirmed and 3663 were marked as candidates.
The archive provided mass estimates for the host stars of 4135 KOIs.  For the
remaining objects, we computed stellar mass estimates on the basis of
$\log{g}$ and stellar radius, when available.
This process yielded a total of 4620 classifiable Kepler objects,
after excluding one object (K07571.01) that was listed with a
planet radius equal to 0.

\begin{figure}[h]
  \includegraphics[width=\columnwidth]{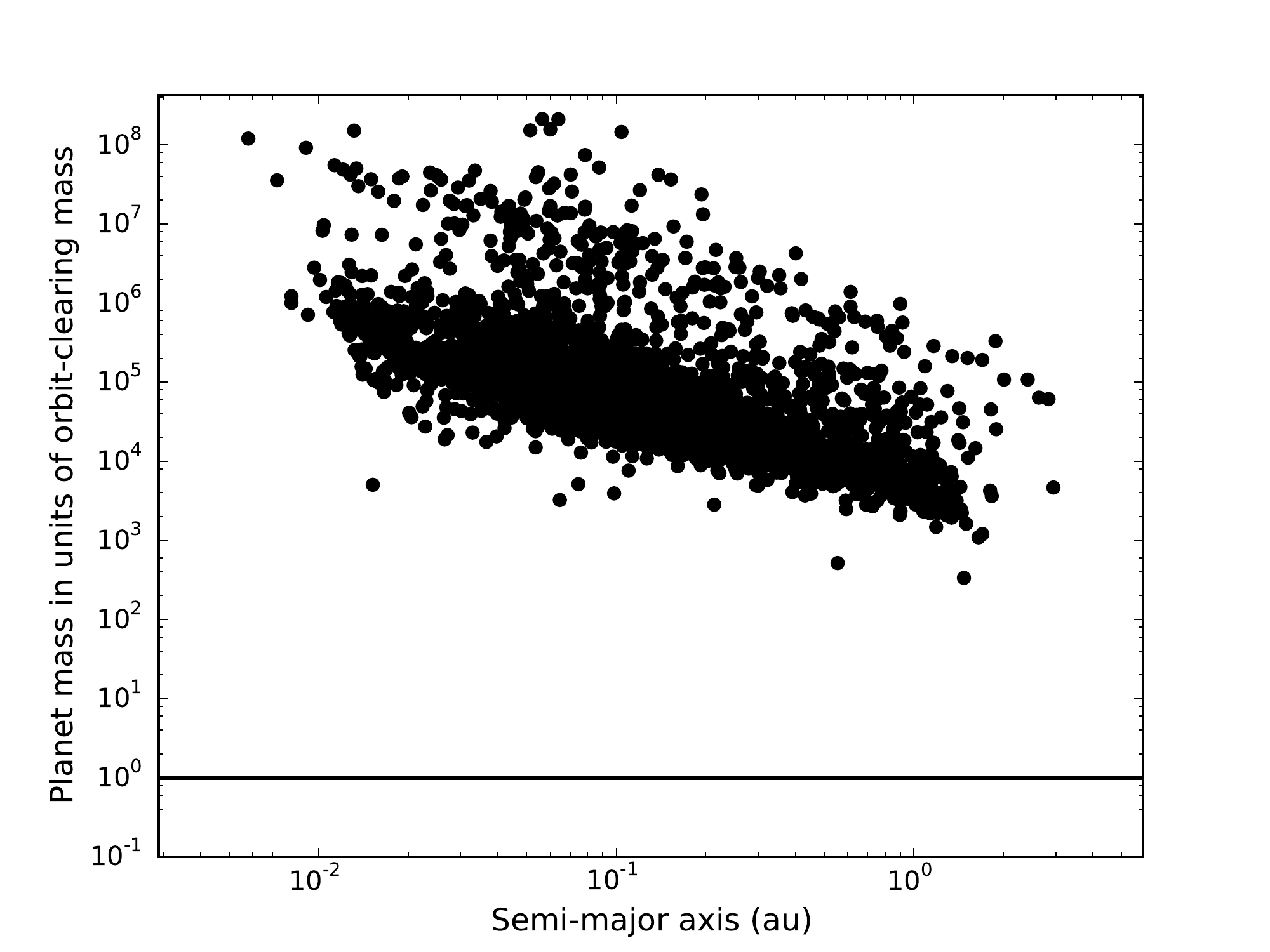}
  \includegraphics[width=\columnwidth]{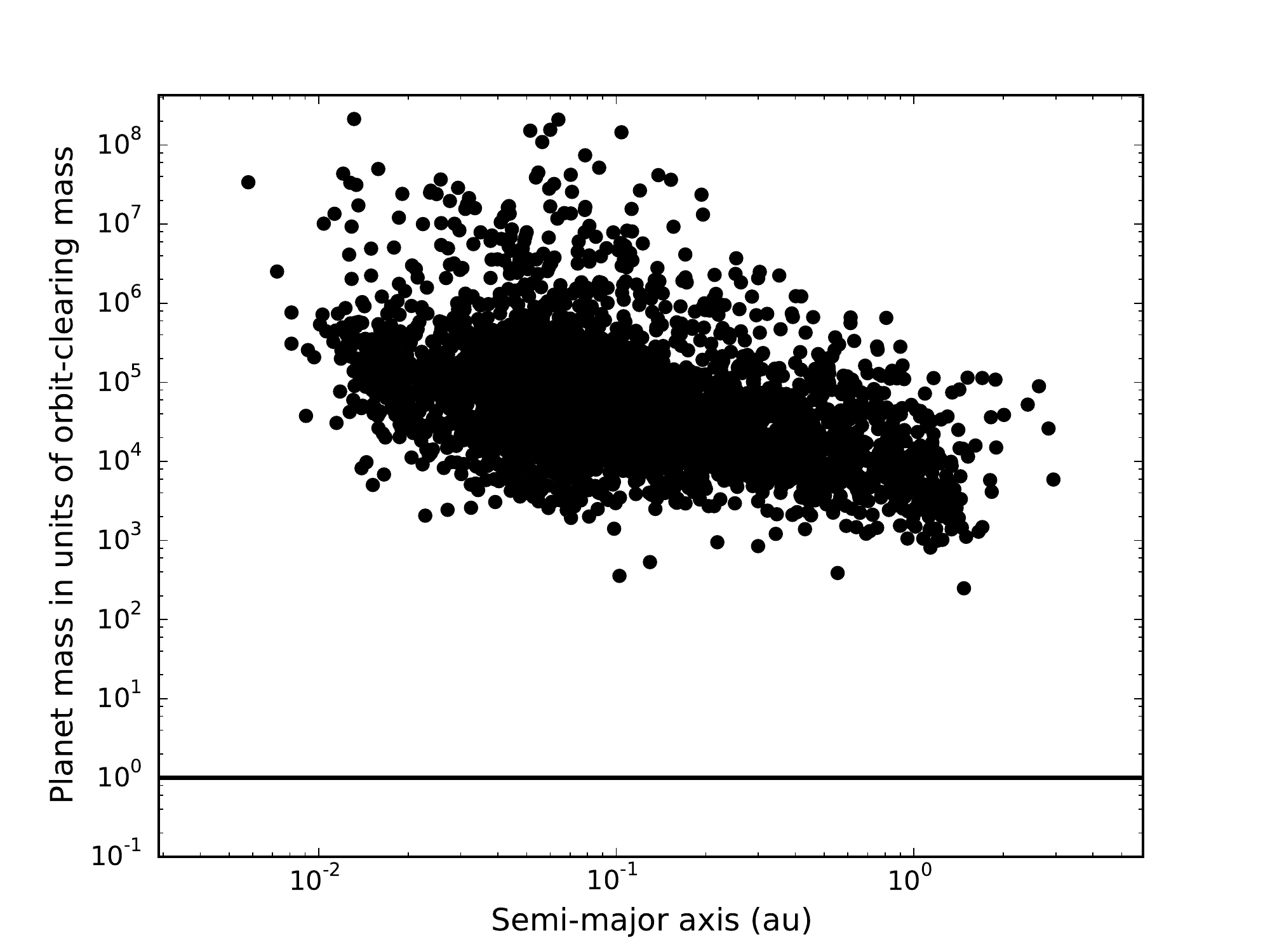}
\caption{Planet test ($C=2\sqrt{3}, t_\star=t_{\rm MS}$) applied to 4276 Kepler
  objects categorized as either ``confirmed'' or ``candidate'' in the
  NASA Exoplanet Archive.
  All objects have a mass that exceeds the mass required to clear the
  corresponding orbital zone over the lifetime of the host star on the
  main sequence ($\Pi \geq 1$).
  Results are robust against the
  choice of the radius-mass relationship.  Two such relationships are
  shown: (top) \citet{fang12exostats} and (bottom) \citet{fabr14}.
}
\label{fig-koi}
\end{figure}

When the planet mass was not available in the archive, 
we applied radius-mass relationships
within their domain of applicability to convert the Kepler
measurements of planet radius to estimates of planet
mass:~\citet[][$R_p<25R_\earth$]{fang12exostats}, \citet[][$R_p<11R_\earth$]{wu13}, \citet[][$R_p<4R_\earth$]{weis14}, \citet{fabr14}, \citet[][$R_p<4R_\earth$]{wolf15}.
We used the
resulting values in equation (\ref{eq-all}) to test whether each KOI
has sufficient mass to clear its orbit.
We found that all KOIs satisfy the criterion irrespective of the
choice of the radius-mass relationship (Figure~\ref{fig-koi}).  For
ease of presentation, we eliminated from the figure one object
(K01174.01) listed with an orbital period of 356 years; its mass
exceeds the orbit-clearing mass by a factor of $\sim 40$.  We also
eliminated 343 objects listed with a planet radius exceeding $20
R_\earth$, leaving a total of 4276 KOIs for display.
A trend of slope $\sim-1$ is visible in the figure because the lower limit
on KOI masses does not vary substantially with $a$ whereas the
orbit-clearing mass scales as $a^{9/8}$.

\subsection{Non-Kepler exoplanets}
As of July 17, 2015, the archive contained 1877 confirmed exoplanets.
We treated
KOIs and five pulsar planets separately (Sections~\ref{sec-koi} and
\ref{sec-pulsar}, respectively).  Eliminating names that include
'Kepler' or 'KOI' and the pulsar planets yielded a sample of 849
objects.
The archive provided mass estimates for the host stars of 696 exoplanets.  For
the remaining objects, we computed stellar mass estimates on the basis
of $\log{g}$ and stellar radius, when available, or from the
exoplanet's orbital period and semi-major axis, when available.  This
process yielded a total of 834 classifiable objects.

\begin{figure}[t]
    \includegraphics[width=\columnwidth]{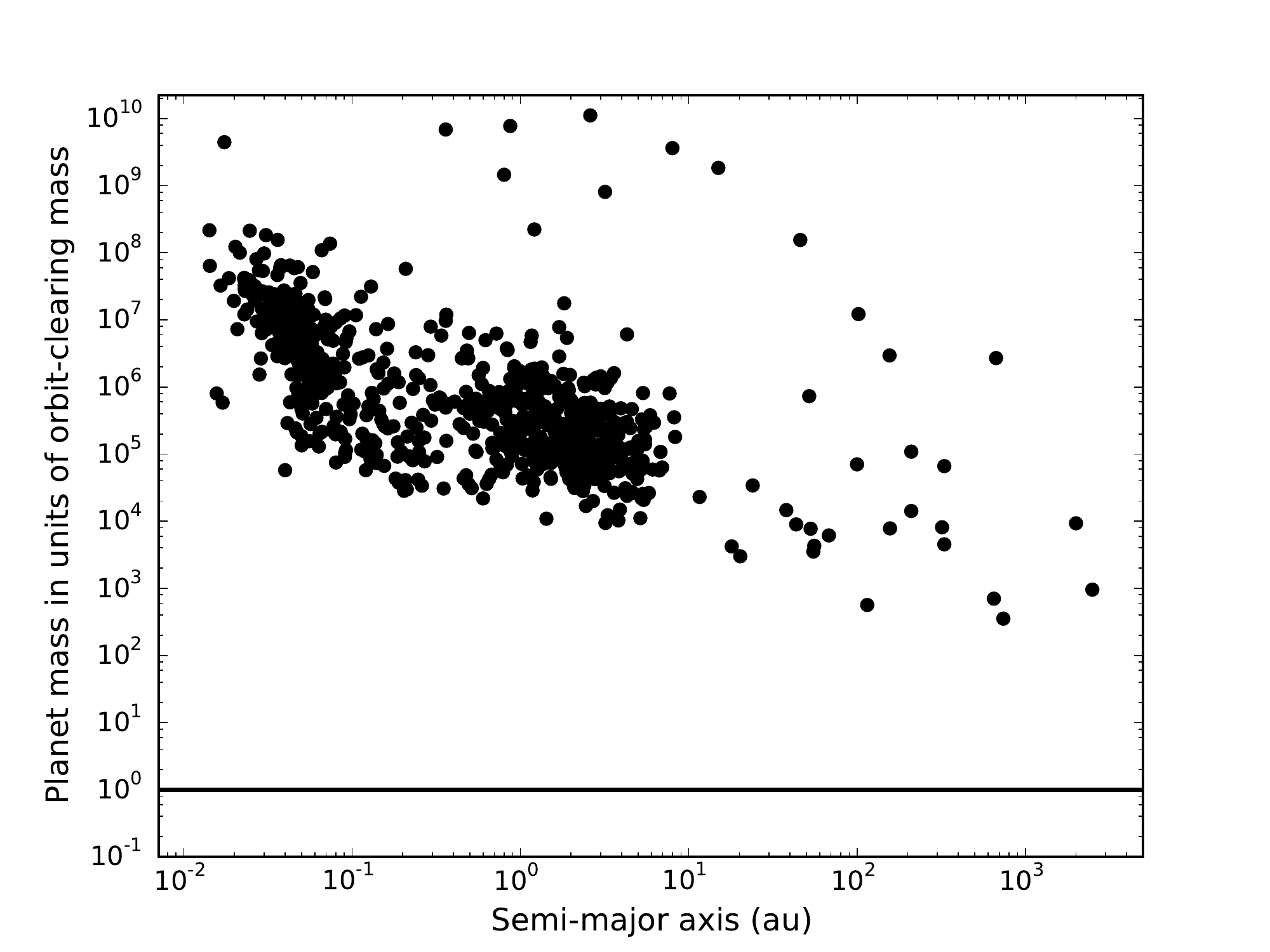}
\caption{Planet test ($C=2\sqrt{3}, t_\star=t_{\rm MS}$) applied to 829
  confirmed, non-Kepler exoplanets in the NASA Exoplanet Archive.  All
  objects have a mass that exceeds the orbit-clearing mass ($\Pi \geq 1$).
\vspace{0.15cm}
}
\label{fig-exo}
\end{figure}

The mass of each exoplanet was obtained from one of three archive
fields: planet mass, $M \sin{i}$, or planet radius, in that order.  We
eliminated 5 exoplanets for which none of these fields were listed,
leaving 829 entries.  When only planet radius was listed, we used a
radius-mass
relationship
to estimate the mass.  The results are robust against the choice of
the radius-mass relationship.  The semi-major axis of each exoplanet
was either available from the archive or computed on the basis of
orbital period and host star mass.
We used the resulting values
in equation (\ref{eq-all}) to test whether non-Kepler exoplanets
have a mass that exceeds the corresponding orbit-clearing mass
(Figure~\ref{fig-exo}).

\clearpage
\subsection{Pulsar planets}
\label{sec-pulsar}

\begin{figure}[h]
      \includegraphics[width=\columnwidth]{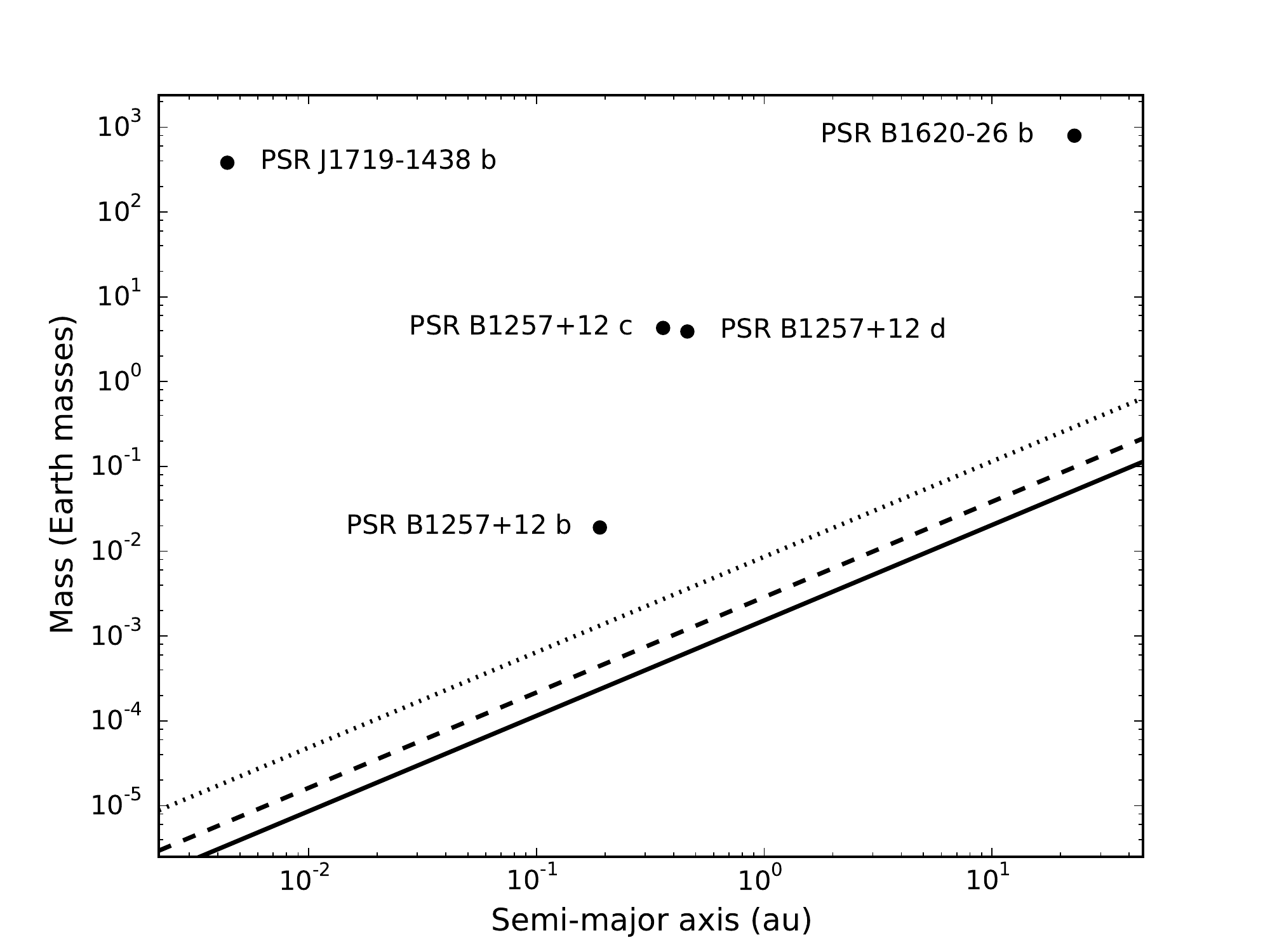}
\caption{Masses of pulsar planets compared to the mass required to
  clear the corresponding orbital zone for a host star of mass 1.4
  $M_\sun$.  Lines show clearing to $2\sqrt{3}$ Hill radii in 10 billion years
  (solid line), 4.3 billion years (dashed line), or 1 billion years
  (dotted line).}

\label{fig-pulsar}
\end{figure}

The choice of the characteristic time scale $t_\star$ for neutron stars is
delicate.  The lifetime of their progenitors on the main
sequence is not directly accessible.  The characteristic age of the
pulsar, while easily measurable from the period and period derivative,
can be very short ($\sim10^4$ y) and not representative of the time
over which planets can clear their orbits.  For ease of
implementation, one could adopt $t_\star = 4.3 \times 10^9\, {\rm y}$
corresponding to the main sequence lifetime for a star of mass 1.4
$M_\sun$, equivalent to the Chandrasekhar limit.  In applying our test
to pulsar planets, we used time scales of $10^9-10^{10} \, {\rm y}$.
As of July 17, 2015, the NASA Exoplanet Archive included 5 pulsar
planets.  We assumed neutron star masses of 1.4 $M_\sun$ and used
equation (\ref{eq-ss}) to test whether all 5 objects have a mass that
exceeds the corresponding orbit-clearing mass
(Figure~\ref{fig-pulsar}).

\section{Maximum Mass}

The 2006 IAU planet definition does not specify an upper limit for the
mass of a planet.  However, the IAU Working Group on Extrasolar
Planets (WGESP), part of the former Division III Planetary Systems
Science, wrote a position statement addressing the need to
differentiate planets from brown dwarfs~\citep{iau07}.  In its
working definition, the WGESP used the limiting mass for thermonuclear
fusion of deuterium, nominally 13 Jupiter masses, to demarcate objects
classified as planets from objects classified as brown dwarfs.
Free-floating objects are also considered in the WGESP working
definition and are never classified as planets.

The WGESP's working definition is based on an observable physical
quantity, which makes classification straightforward.  The mass of the
object to be classified is compared to the limiting mass for deuterium
fusion.  This criterion does not require that the object actually
experienced deuterium fusion.  This distinction is critical because it
is far more practical to measure the mass of an object than to understand
its evolutionary history.  In doing so, the WGESP acknowledged that it
was preferable to risk a small fraction of inaccurate classifications
(e.g., classifying an object that experienced deuterium fusion at some
point in its history as a planet) than to build a classification
around the detection of a signature that is not reliably observable.
The criterion proposed in this paper is very much aligned with the
WGESP's philosophy.  The mass of the object to be classified is
compared to the corresponding orbit-clearing mass via the planet
discriminant $\Pi$.  The criterion does not require that the object
actually cleared its orbit.

\section{Astrophysics of planet formation}
In accordance with the IAU's and WGESP's approaches, we purposefully
developed a taxonomic tool that is not based on hypotheses related to
planet formation.  We chose not to allow our incomplete and possibly
incorrect understanding of the planet formation process to interfere
with the design of a planet discriminant.  Concordance of the
classification scheme with formation and evolution processes may
ultimately become a desirable trait, but it is only one of several
desirable traits of a good taxonomy.

In spite of our agnosticism, the planet discriminant appears to
identify actual physics of the planet formation process.  For instance,
there is a gap of more than 3 orders of magnitude in the value of
$\Pi$ between planets and dwarf planets in the solar system.  This gap
may illuminate the physics of accretion time scales, oligarchic
growth, and dynamical evolution of the solar
system~\citep[e.g.,][]{armi13}.  Likewise, all classifiable exoplanets
appear to have $\Pi$ values well above 1.  Although this
characteristic may currently be due to observational selection
effects, persistence of this trait in future surveys with more
sensitive instruments would compel us to seek an explanation.

\section{Discussion}
\label{sec-discuss}

\subsection{Extent of orbital zone to be cleared}
We adopted the size of a planet's feeding zone, corresponding to
$C=2\sqrt{3}$, as the minimum extent of the orbital zone to be
cleared.  This theoretical value is in agreement with the results of
numerical simulations.  \citet{morr15} described the
boundaries of orbital zones that are cleared over a wide range of
planet-to-star mass ratios (10$^{-9}$--10$^{-1.5}$) and planet radii
(0.001 $R_H$--0.1 $R_H$) in the context of the planar, circular,
restricted three-body problem.  The boundaries remain within a factor
of 0.6--1.5 of $2\sqrt{3} R_H$ over this entire range of conditions.
Similar extents may apply to orbits with eccentricities up to about
0.3~\citep{quil06faber}.  In non-planar systems of multiple
interacting planets, the extent of the zone that is cleared may of
course exceed $2\sqrt{3} R_H$ because of additional perturbations.

We also evaluated the planet discriminant for a slightly larger
extent, corresponding to $C=5$, which roughly matches
half the observed separations between tightly packed exoplanets.  Most
planets in Kepler multi-planet systems are separated by at least 10
mutual Hill radii\footnote{The mutual Hill radius for two planets has
  the same form as equation (\ref{eq-rh}) but involves the sum of the
  planetary masses and the average of the semi-major axes.} with a
mean spacing of $\sim 20$ mutual Hill
radii~\citep{fang13exopps,liss14}.  The value of 10 mutual Hill radii
also represents the approximate minimum spacing required for long-term
dynamical stability of planets on circular, coplanar orbits in
multi-planet systems~\citep[][and references therein]{cham96,puwu15}.

It is noteworthy that the orbit-clearing criterion does not depend
strongly on the adopted value of $C$; other values besides $2\sqrt{3}$
and 5 may be considered.

\subsection{Characteristic time scale}
We adopted the lifetime of the host star on the main sequence as a
sensible time scale, but other choices can be readily
accommodated. Restricting the time scale to 10\% of the stellar
lifetime, for instance, would increase the orbit-clearing mass by a
factor of $\sim6$.  Doing so would not change the classification of
any of the objects considered here.

\subsection{Clearing process}
Numerical simulations in the context of the planar, circular,
restricted three-body problem can reveal the fate of particles that
orbit in the vicinity of a planet~\citep{morr15}.  For small values of
the planet-to-star mass ratio, simulations show that removal by
collisions with the planet is more frequent than removal by
scattering, although scattering remains dominant in the limit of small
planetary radii.  In addition, removal times are shorter when
collisions are the dominant clearing mechanism~\citep{morr15}.  An
object with $\Pi \geq 1$ can therefore clear its orbital zone within
the prescribed time scale, whether collisions or scattering events
prevail in the clearing process.  We prefer the simplicity of a single
lower bound provided by the diffusive time scale over the construction
of a hybrid criterion with both collisional and diffusive time
scales~\citep[e.g.,][]{levi08}.

\subsection{Bodies on eccentric orbits}
Although the orbit-clearing criterion was developed for
planets on circular orbits, the essence of the metric is based on a
random walk in orbital energy.  Gravity kicks will modify the orbits
of small bodies regardless of the planet's eccentricity, such that the
basic concept of orbit clearing remains applicable.  How the
orbit-clearing time scale varies as a function of orbital eccentricity
is an interesting question left for future work.  Because we
anticipate that the criterion will hold to first order and because we
favor ease of implementation, we suggest applying the planet
discriminant $\Pi$ regardless of orbital eccentricity.

\subsection{Co-orbitals}
A planet can clear its orbit
yet capture small bodies in tadpole, horseshoe, and quasi-satellite
orbits near the Lagrange equilibrium points~\citep{murr99}.
Well-known examples include the Trojan asteroids associated with
Jupiter. Our proposed criterion follows the IAU definition in that the
existence of bodies in such orbits has no bearing on the
classification.  More exotic configurations can be handled by the
proposed criterion as well.  Co-orbital planets have not been
discovered to date, but some configurations are in principle stable
over long periods of time~\citep{salo88,smit10}.  For instance, one
could differentiate between co-orbital planets (where the planet
masses individually exceed $M_{\rm clear}$) and a planet-trojan system
(where only one of the bodies meets the criterion).

\subsection{Satellites}
The IAU has not formally defined satellites, which are informally
understood to be celestial bodies that orbit planets, dwarf planets,
or asteroids.  Satellites to planets will have little or no impact on
the classification if the satellite-to-planet mass ratio is low.  At
higher values of the mass ratio, satellites may affect the
classification because it is the sum of the component masses in a
bound system that determines the ability to clear an orbital zone.
The terminology could in principle differentiate between two-body
planets (where the sum of the masses exceeds $M_{\rm clear}$, but the
individual component masses do not) and double planets (where the
individual masses both exceed $M_{\rm clear}$).

Improvements to the classification are needed to deal with celestial
bodies that orbit brown dwarfs.  Because such bodies do not orbit a
star or stellar remnant, we do not consider them planets.

\subsection{Circumbinary planets}
Celestial bodies in orbit around a system of bound stars can be
classified with the proposed criterion by using the sum of the stellar
masses in equation (\ref{eq-ss}) and a time scale corresponding to the
shortest
stellar lifetime.

\subsection{Free-floating objects}
In conformity with the 2006 IAU planet definition and WGESP
recommendations, we do not consider planetary objects that never
orbited a star or no longer orbit a star (i.e., free-floating objects)
to be planets.  Some scientists dislike the concept of a planet
definition that depends on context and would prefer to focus on
intrinsic properties.  However, there are instances in which context
justifiably prevails in the classification (e.g., asteroid
vs.\ meteorite, magma vs.\ lava, cloud vs.\ fog), and there is no
reason to dispense with a useful distinction in the taxonomy of
planets.

\subsection{Migration and scattering}
An object that is classified as a planet, with $\Pi \geq 1$, will lose
this classification if it migrates or is scattered to a distance from
its host star where $\Pi < 1$.  This reclassification is similar to
the reclassification of asteroids as meteorites upon impact with a
planet or moon.  According to the proposed criterion, an Earth-mass
body orbiting a solar-mass star at 400 au, where $\Pi < 1$, would not
be considered a planet, whether it formed there or was transported
there after formation.

\subsection{Advantage over other proposed metrics}
\citet{sote06} proposed a planet discriminant that requires the mass
of a body to be 100 times the mass of all other bodies that share its
orbital zone.  \citet{levi08} favored a definition in which an object
that is part of a smooth size distribution is not a
planet. \citet{vals09} proposed a criterion in which the mass of a
body must exceed the mass of all bodies that come close to or cross
its path by a factor of 1000.  The difficulty with implementing these
criteria is that the mass or size distribution of the neighboring
small bodies must be measured or estimated.  In other words, it is not
possible to classify a celestial body until knowledge about
neighboring bodies is secured.  It would, therefore, be difficult to
classify most exoplanets based on these definitions.  The main
advantage of the orbit-clearing criterion proposed here is that no
such knowledge is required.

\section{On Roundness}
Equation (\ref{eq-all}) provides a quantifiable criterion that
addresses the first and third aspects of the 2006 IAU planet
definition.  A separate issue is whether the second requirement, i.e.,
roundness, is necessary.  It is possible that every object that can
clear its orbital zone is ``nearly round,''
which would make the roundness requirement superfluous.

\citet{tanc08} examined the size and density bounds that guarantee
approximate roundness in planetary bodies.  They recommended a
diameter threshold of 800 km for rocky bodies, which corresponds to a
mass threshold of $M_{\rm round} \sim 10^{-4} M_\earth$ for a density
of 2.5 gcm$^{-3}$.  \citet{line10} find a slightly smaller
diameter threshold of 600 km.

A planet that has cleared its orbital zone is expected to have
accumulated a mass on the order of the isolation mass, the mass of the
planetesimals in its feeding zone.  The mass of the planet may be
substantially larger than the isolation mass if there has been
migration and resupply of disk material, migration of the planet
through the disk, planet mergers, post-formation accretion of
asteroidal material, or a combination of these factors.

In convenient units, \citet{armi13}'s expression for the isolation mass
reads
\begin{equation}
  \frac{M_{\rm iso}}{M_\earth} = 6.6 \times 10^{-2} \left(\frac{M_\star}{M_\sun}\right)^{-1/2} \left(\frac{\Sigma_p}{10\, {\rm gcm}^{-2}}\right)^{3/2} \left(\frac{a_p}{1\, {\rm au}}\right)^{3},
\label{eq-iso}
\end{equation}
where $\Sigma_p$ is the local surface density of the planetesimals.
The functional form of the surface density is uncertain, but a common
model is
\begin{equation}
\Sigma_p = \Sigma_0 \left(\frac{a_p}{1\, {\rm au}}\right)^{-3/2},
\end{equation}
which yields an isolation mass
\begin{equation}
  \frac{M_{\rm iso}}{M_\earth} = 6.6 \times 10^{-2} \left(\frac{M_\star}{M_\sun}\right)^{-1/2} \left(\frac{\Sigma_0}{10\, {\rm gcm}^{-2}}\right)^{3/2} \left(\frac{a_p}{1\, {\rm au}}\right)^{3/4}.
\label{eq-isosigma}
\end{equation}
Evaluation of this expression over a wide range of conditions (1 $<
\Sigma_0 <$ 10 gcm$^{-2}$, 0.01 $< a_p <$ 100 au) shows that 
\begin{equation}
M_{\rm iso} > M_{\rm round}.
\end{equation}
This approximate calculation suggests that
every object that has cleared its orbital zone and accumulated at
least an isolation mass
worth of material is nearly round.  Because of residual uncertainties
related to the size at which planetary bodies become round, the exact
surface density profile of planetesimals, and the process of planet
formation, it is difficult to gauge roundness on theoretical grounds
with greater certainty.  However, attempting to gauge roundness
observationally would be equally difficult and lead to comparable
uncertainties.  The threshold for roundness depends on the interior
composition of the body and temperature-dependent material
strength~\citep{tanc08}, which are not observable from Earth.

\vspace{0.15cm}

\section{Possible improvements to the IAU planet definition}

Because a quantitative orbit-clearing criterion can be applied to all
planets and exoplanets, it is possible to extend the 2006 IAU planet
definition to stars other than the Sun and to remove any possible
ambiguity about what it means to clear an orbital zone.

In addition, because it is probable that all objects that satisfy the
orbit-clearing criterion also satisfy the roundness criterion, it is
possible to simplify the definition by removing the latter criterion.

One possible formulation (with $C=2\sqrt{3}, t_\star = t_{\rm MS}$) is as follows:
\begin{quote}
A planet is a celestial body that (a) is in orbit around one or more
stars or stellar remnants, (b) has sufficient mass to clear [or
  dynamically dominate] the neighbourhood around its orbit, i.e., $\Pi
\geq 1$, (c) has a mass below 13 Jupiter masses, a nominal value close to 
the limiting mass for thermonuclear fusion of deuterium.

For single-star systems, $\Pi \geq 1$ when 
\begin{equation*}
\frac{M_p}{M_\earth} \gtrsim 1.2 \times 10^{-3} \left(\frac{M_\star}{M_\sun}\right)^{5/2} \left(\frac{a_p}{1\, {\rm au}}\right)^{9/8},
\end{equation*}
where $M$ is mass, $a$ is semi-major axis, and subscripts $p, \star, \earth, \sun$ refer to the planet, star, Earth, and Sun, respectively.
\end{quote}

\section{Acknowledgments}

Steven Soter, Yanqin Wu, Brad Hansen, Luke Dones, Julio Fern\'andez,
Sarah Morrison, Renu Malhotra, Adam H.\ Greenberg, M.\ Oliver Bowman,
Gerald McKeegan, Daniel Fischer, and an anonymous reviewer provided
useful comments on the manuscript.  We also benefited from
conversations with Hal Levison, Nader Haghighipour, Giovanni
Valsecchi, Gonzalo Tancredi, Didier Queloz, Andreas Quirrenbach, Eric
Mamajek, David Kipping, Paul Wilson, Charles Lineweaver, and Kevin
McKeegan.  This research has made use of the NASA Exoplanet Archive,
which is operated by the California Institute of Technology, under
contract with the National Aeronautics and Space Administration under
the Exoplanet Exploration Program.

\bibliography{bib}
\end{document}